\documentclass[iop, twocolumn, tighten]{emulateapj}
\usepackage{amsmath}

\newcommand{\eR}{{\boldsymbol e}_R}
\newcommand{\eRd}{{\boldsymbol e}_R^\prime}
\newcommand{\eO}{{\boldsymbol e}_\mathrm{O}}
\newcommand{\eS}{{\boldsymbol e}_\mathrm{S}}
\newcommand{\espin}{{\boldsymbol e}_\mathrm{spin}}
\newcommand{\emax}{{\boldsymbol e}_\mathrm{M}}
\newcommand{\phimax}{\hat{\phi}_\mathrm{M}}
\newcommand{\thetamax}{\hat{\theta}_\mathrm{M}}

\newcommand{\fspin}{\mathit{f}_\mathrm{spin}}
\newcommand{\forb}{\mathit{f}_\mathrm{orb}}
\newcommand{\fobs}{\mathit{f}_\mathrm{obs}}
\newcommand{\fif}{\mathit{f}}

\newcommand{\sphi}{\sin{\phi}}
\newcommand{\cphi}{\cos{\phi}}
\newcommand{\szeta}{\sin{\zeta}}
\newcommand{\czeta}{\cos{\zeta}}
\newcommand{\stheta}{\sin{\theta}}
\newcommand{\ctheta}{\cos{\theta}}
\newcommand{\si}{\sin{i}}
\newcommand{\ci}{\cos{i}}

\newcommand{\Porb}{P_\mathrm{orb}}
\newcommand{\Pspin}{P_\mathrm{spin}}

\newcommand{\ThetaM}{\Theta_\mathrm{eq}}
\newcommand{\tThetaM}{\tilde{\Theta}_\mathrm{eq}}

\newcommand{\trans}{\mathsf{T}}

\slugcomment{}
\shortauthors{Kawahara}
\shorttitle{FM of directly imaged exoplanets}
\begin{document}
\title{Frequency Modulation of Directly Imaged Exoplanets: Geometric Effect as a Probe of Planetary Obliquity}

\author{Hajime Kawahara\altaffilmark{1,2}} 
\altaffiltext{1}{Department of Earth and Planetary Science, 
The University of Tokyo, Tokyo 113-0033, Japan}
\altaffiltext{2}{Research Center for the Early Universe, 
School of Science, The University of Tokyo, Tokyo 113-0033, Japan}
\email{Electronic address: kawahara@eps.s.u-tokyo.ac.jp}

\begin{abstract}
We consider the time-frequency analysis of a scattered light curve of a directly imaged exoplanet. We show that the geometric effect due to planetary obliquity and orbital inclination induce the frequency modulation of the apparent diurnal periodicity. We construct a model of the frequency modulation and compare it with the instantaneous frequency extracted from the pseudo-Wigner distribution of simulated light curves of a cloudless Earth. The model provides good agreement with the simulated modulation factor, even for the light curve with Gaussian noise comparable to the signal. Notably, the shape of the instantaneous frequency is sensitive to the difference between the prograde, retrograde, and pole-on spin rotations. While our technique requires the albedo map to be static, it does not need to solve the albedo map of the planet. The time-frequency analysis is complementary to other methods which utilize the amplitude modulation. This paper demonstrates the importance of the frequency domain of the photometric variability for the characterization of directly imaged exoplanets in future research.
\end{abstract}


\section{Introduction}

The photometric variability of scattered light is expected to be an important probe for the characterization of directly imaged exoplanets in the near future. In the context of the habitable planet search, \citet{2001Natur.412..885F} demonstrated that inhomogeneous clouds and the surface components of Earth generate photometric variability due to a spin rotation. Surface distribution inversion techniques have been extensively studied \citep[e.g.][]{2009ApJ...700..915C,2009ApJ...700.1428O,2010ApJ...715..866F,2010ApJ...720.1333K,2011ApJ...731...76C,2011ApJ...738..184F,2011ApJ...739L..62K,2012ApJ...755..101F}. 

The inversion method using the diurnal and annual variability, referred to as the spin-orbit tomography, retrieves not only the two-dimensional surface distribution but also the planet obliquity \citep{2010ApJ...720.1333K, 2011ApJ...739L..62K,2012ApJ...755..101F}. Planet obliquity (axial tilt) is an important parameter for an exoplanet's environment \citep[e.g.][]{1997Icar..129..254W,2003IJAsB...2....1W} and formation theory \citep[e.g.][]{1999Icar..142..219A,2001Icar..152..205C,2007ApJ...671.2082K} but it has not yet been measured. Several observational features aside from the photometric variability have been proposed to probe obliquity, including the difference between the ingress and egress shapes in a transit curve \citep{2002ApJ...574.1004S,2003ApJ...588..545B,2010ApJ...709.1219C}, the modulation of the planet's radial velocity\citep{2012ApJ...760L..13K}, and the Rossiter-McLaurin effect at the planet's occultation \citep{2015ApJ...808...57N}. 

In the framework of spin-orbit tomography, the obliquity is simultaneously derived from the retrieval of the surface map \citep{2010ApJ...720.1333K, 2011ApJ...739L..62K,2012ApJ...755..101F}. The generality of obliquity retrieval from photometric variation was recently studied by \citet{2015arXiv151105152S}. They explained how the obliquity affected the shape and location of the kernel of the scattered light. In essence, these methods use information on the amplitude modulation to extract spin information. Hence, the albedo map of the planet contributes nuisance parameters to the global light curve fit. In this paper, we focus on the frequency domain of the photometric variability, rather than the amplitude modulation. We utilize the frequency modulation as an estimator of the obliquity, which is less sensitive to the albedo distribution of the planet.

Regarding photometric variability in the frequency domain, \citet{2008ApJ...676.1319P} presented a pioneering work applying the autocorrelation function to the simulated photometric variation. They showed that the photometric variation of the simulated Earth contains sufficient information to measure the spin rotation period despite variable weather patterns. \citet{2015A&A...579A..21V} studied the Fourier coefficient of the photometric variation for various surface types of Earth-like planets and several configurations (i.e. obliquity and inclination). The Fourier analysis of the reflected light curve has also been used in the asteroid field \citep[e.g.][]{1906ApJ....24....1R,1989Icar...78..311B}. We extend the frequency analysis of the photometric variability to the time-frequency analysis to consider the frequency modulation. We show that both the orientation of the spin axis and the orbital inclination modulate the frequency of the apparent variability of the scattered light. In this paper, we concentrate on the geometric effect of the frequency modulation and restrict the scope of targets to the light scattered by the static planetary surface. 

The rest of the paper is organized as follows. In Section 2, we construct a simple model of the frequency modulation due to geometric effects. In Section 3, we simulate photometric variability assuming the Lambert model of static albedo distribution. We extract the instantaneous frequency from the pseudo-Wigner distribution of the simulated light curve. We discuss the other effects that cause the frequency modulation in Section 4. We also compare our results with the spin-orbit tomography. In Section 5, we summarize our findings.  

\section{Apparent Diurnal Periodicity}

Let us first consider the prograde rotation of an aligned planet with a static surface distribution. The spin rotation frequency is defined by $\fspin =  1/\Pspin$, where $\Pspin$ is the sidereal day of the planet (the spin rotation period). The sidereal day is the period required for the planet to make one rotation about the inertial reference frame of the stellar system. The period of the photometric variation is identical to the so-called synodic day, which is defined as the period that it takes to rotate once around the central star. The apparent diurnal periodicity derived from the photometric variability is then $\fobs = \fspin - \forb$, where $\forb = 1/\Porb$ is the orbital frequency and $\Porb$ is the orbital period. For the retrograde rotation (the obliquity of $180^\circ$), we obtain $\fobs = \fspin + \forb$. We are then faced with the question of how the obliquity changes the apparent diurnal periodicity. 

\begin{figure}[htbp]
\begin{center}
\includegraphics[width=\linewidth]{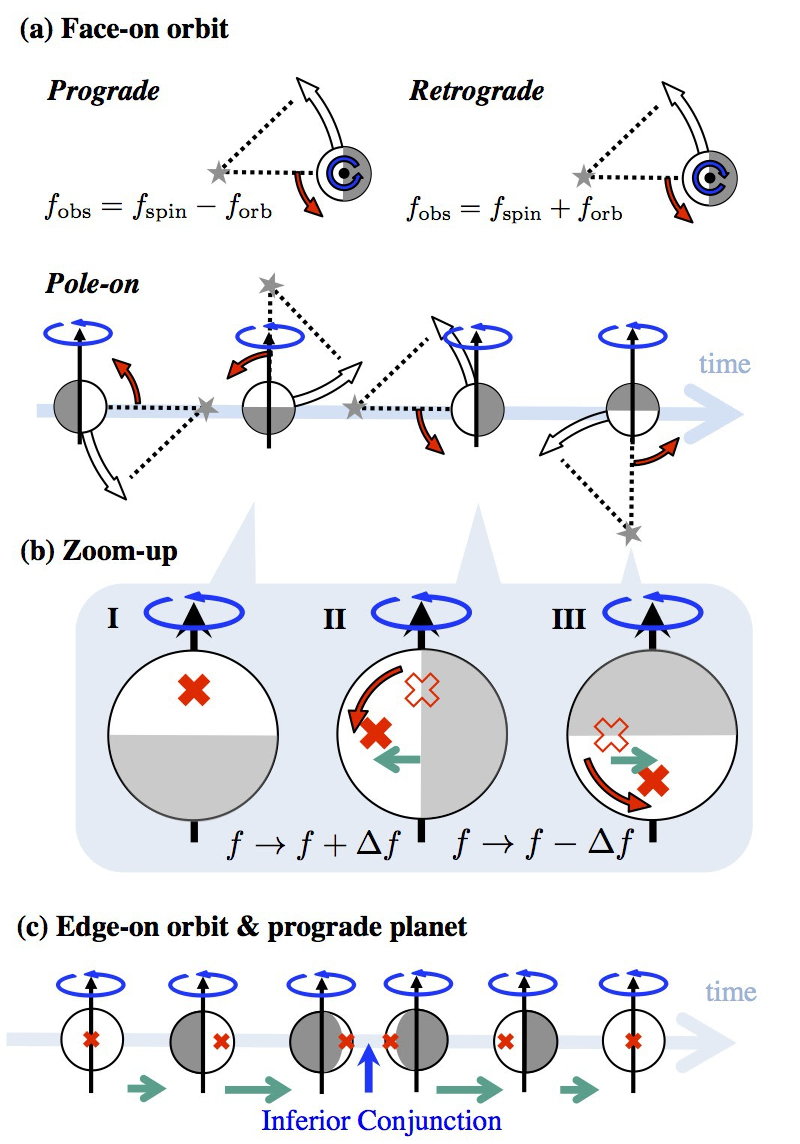}
\caption{Schematic explanation of the geometric frequency modulation. The star indicates the position of the central star. The black and white arrows are the spin axis vector of the planet and the direction of orbital motion. The red arrows indicate the direction of the shift of the illuminated area shown in white. The configurations for the prograde, retrograde, and pole-on rotations are shown in panel (a). Panel (b) shows an enlarged view of the pole-on case. The filled cross shows the weighted center of the scattered light. According to the orbital motion, the longitude of the weighed center moves on the planet's disk as indicated by the green arrow. The apparent rotation is accelerated or decelerated by the motion of the weighted center. Panel (c) explains the frequency modulation induced by the orbital inclination. See the text for the details. \label{fig:obliquity_type} }
\end{center}
\end{figure}

Figure \ref{fig:obliquity_type} illustrates how the obliquity and orbital inclination induce frequency modulation. Panel (a) displays prograde, retrograde, and pole-on planets in a face-on orbit. The illuminated area painted in white moves in the direction of the red arrow. Because the rotation axis of the illuminated area is aligned with the spin axis, the photometric frequency of the prograde and retrograde planets is stationarily shifted as $\fobs = \fspin \pm \forb$. For the case of pole-on planets, the rotation axis of the illuminated area is not aligned with the spin axis. In panel (b), the representative point (the weighted center) of the illuminated area is indicated by a filled cross. When the latitude of the weighted center moves from the center (I) to the left (II), the apparent rotation speeds up. The shift of the weighted center to the right decelerates the apparent rotation (III). Thus, the motion of the illuminated area modulates the frequency of the apparent rotation.

The motion of the illuminated area due to orbital inclination also induces frequency modulation according to the same principle. As shown in panel (c), the shift of the illuminated area around the inferior conjunction negatively modulates the apparent rotation rate. In general, frequency modulation due to orbital inclination is significant near the inferior conjunction. We call these types of frequency modulation the geometric frequency modulation.

\subsection{Maximum Weighted Longitude Approximation of the Synodic Diurnal Modulation}

\begin{figure*}[htbp]
\begin{center}
\includegraphics[width=0.32\linewidth]{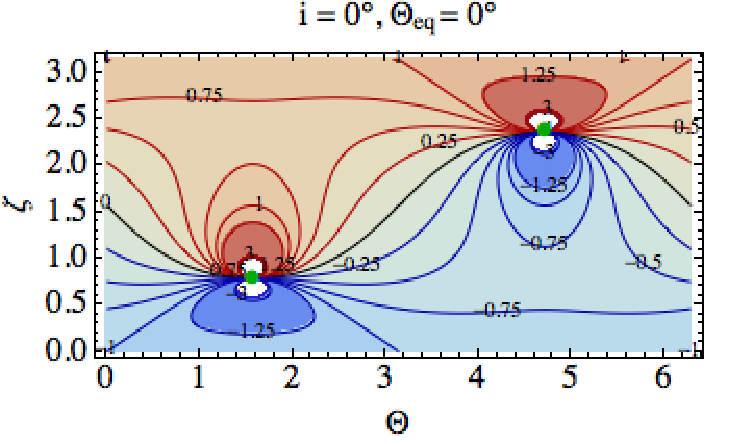}
\includegraphics[width=0.32\linewidth]{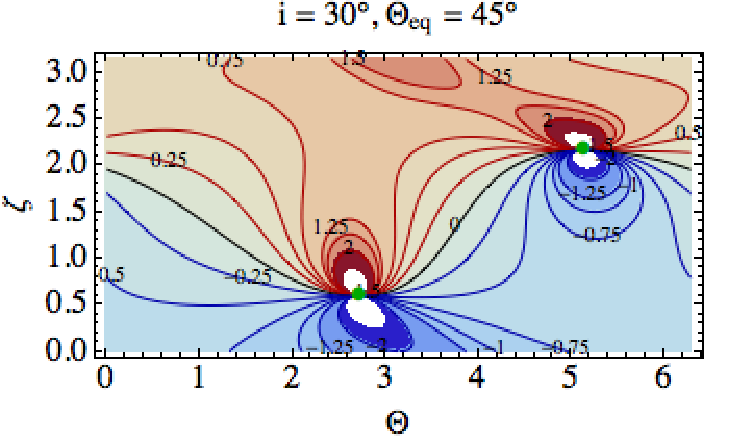}
\includegraphics[width=0.32\linewidth]{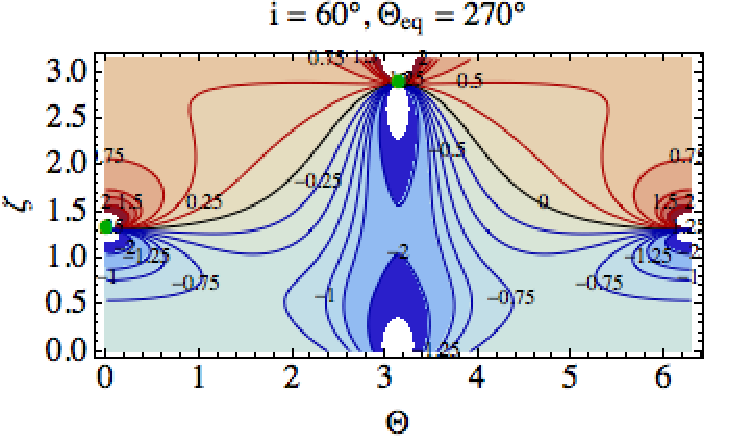}
\caption{\label{fig:modfac} Examples of the modulation factor under the maximum weighted longitude approximation, $\epsilon_\zeta(\Theta)$. We assume $(i=0^\circ, \ThetaM=0^\circ)$, $(i=30^\circ, \ThetaM=45^\circ)$, and $(i=60^\circ, \ThetaM=270^\circ)$ from left to right. The singular points are marked by the green points. }
\end{center}
\end{figure*}

The apparent periodicity depends on both the spin vector and the albedo distribution. The integrated light from the Lambert surface is expressed as \citep{2010ApJ...720.1333K}
\begin{eqnarray}
\label{eq:lambert}
I_b \propto \int_{S} \, a(\phi, \theta) W_V (\phi, \theta) W_I (\phi, \theta)  \sin{\theta} d \theta d \phi,  
\end{eqnarray}
where $a(\phi, \theta)$ is the albedo distribution on the planetary surface of ($\phi, \theta$), $S$ is the illuminated and visible area, $W_V(\phi, \theta) = \eO \cdot \eR$ and $W_I(\phi, \theta) = \eS \cdot \eR$ are the weight functions of the visible and illuminated areas, and $\eS, \eO$ and $\eR$ are the unit vectors of $\overrightarrow{\mathrm{ps}}$, $\overrightarrow{\mathrm{po}}$  (p=the planet center, s=the stellar center, and o=the observer) and, the normal vector of the surface. 

It is useful to exclude the influence of the albedo distribution from the model. Assuming that variability can be mainly attributed to the surface at the maximum weighted longitude, we construct a model of photometric periodicity by computing the phase shift of the maximum weighted longitude, $\phimax$.  If the spin axis is aligned with the orbital axis, i.e., if the planetary obliquity $\zeta$ is zero, then $\phimax$ moves according to the spin rotation. Then, we obtain $\phimax = - \Phi$, where $\Phi$ is the diurnal phase of the spin rotation defined by $\Phi = \fspin t/2 \pi$. The minus sign implies that the Sun rises from the east and sets in the west.

The orbital motion modulates $\phimax$ for the non-zero obliquity. The instantaneous frequency of the periodicity at the maximum weighted longitude is given by 
\begin{eqnarray}
\label{eq:insta}
\fobs &=& - \frac{1}{2 \pi} \frac{\partial \phimax}{\partial t} = - \frac{\partial \phimax}{\partial \Theta} \forb, 
\end{eqnarray}
where $\Theta = \forb t/2 \pi$ is the orbital phase. Using the relation 
\begin{eqnarray}
\label{eq:maxaxa}
\frac{\partial (\phimax + \Phi) }{\partial \Theta} &=& \frac{\partial \phimax }{\partial \Theta} + \frac{\partial \Phi }{\partial \Theta} \nonumber = \frac{\partial \phimax }{\partial \Theta} + \frac{\fspin}{\forb}
\end{eqnarray}
we rewrite equation (\ref{eq:insta}) as 
\begin{eqnarray}
\label{eq:maxf}
\fobs = \displaystyle{\fspin + \epsilon_\zeta(\Theta) \, \forb}, 
\end{eqnarray}
where we define the modulation factor as
\begin{eqnarray}
\label{eq:maxfd}
\epsilon_\zeta(\Theta) &\equiv& - \frac{\partial (\phimax + \Phi) }{\partial \Theta} = - \frac{\kappa^\prime (\Theta)}{1 + \kappa(\Theta)^2},\\
\kappa(\Theta) &\equiv& \tan{(\phimax+\Phi)}.
\end{eqnarray}

 To compute $\kappa(\Theta)$, we use the inertia coordinate system described in \cite{2012ApJ...755..101F}:
\begin{eqnarray}
  \label{eq:eS}
\eS &=& (\cos{(\Theta-\ThetaM)}, \sin{(\Theta-\ThetaM)}, 0)^\trans,\\
  \label{eq:eO}
\eO &=& (\si \cos{\ThetaM}, - \si \sin{\ThetaM}, \ci)^\trans, 
\end{eqnarray}
where  $\ThetaM$ is the orbital phase at the equinox. Because $W_I$ and $W_V$ are the inner products, $\eO \cdot \eR$ and $\eS \cdot \eR$, $\eR$ has maximum weight when $\eO \cdot \eR = \eS \cdot \eR$. Then, we obtain the vector from the center to the maximum weighted point as
\begin{eqnarray}
&\,& \emax = \frac{\eS+\eO}{|\eS+\eO|} \\
&=& \frac{1}{L} \left(
\begin{array}{c}
\cos{(\Theta-\ThetaM)} + \cos{\ThetaM} \sin{i} \\
\sin{(\Theta-\ThetaM)} - \sin{\ThetaM} \sin{i} \\
\cos{i}
\end{array} \right),
\end{eqnarray}
where $L \equiv |\eS+\eO|=\sqrt{2+2\cos{\Theta}\sin{i}}$. 

We set the Cartesian coordinates fixed on the planetary surface $\eRd(\phi,\theta) = (\cphi \stheta, \, \sphi \stheta, \, \ctheta)^\trans $. We call this the surface Cartesian coordinate. The spin vector is $\espin^\prime =(0,0,1)^\trans$ in the surface Cartesian coordinate. The conversion between the inertia coordinate $\eR$ and the surface Cartesian coordinate $\eRd$ is expressed as 
\begin{eqnarray}
  \label{eq:eR}
\eR &=&  R (\zeta) \, \hat{S}(\Phi) \, \eRd (\phi, \theta) \nonumber \\
&=& R (\zeta) \, \eRd (\phi+\Phi, \theta) \nonumber \\
&=& \left(
\begin{array}{c}
\cos{(\phi+\Phi)} \stheta \\
\czeta \sin{(\phi+\Phi)} \stheta +  \szeta  \ctheta \\
- \szeta \sin{(\phi+\Phi)} \stheta + \czeta \ctheta  
\end{array} \right),
\end{eqnarray}
where $R(\zeta)$ is the rotation matrix for the clockwise rotation around the $x$-axis and $\hat{S}(\Phi)$ is a rotation operator of $\phi \to \phi + \Phi$. To describe $\emax$ in the surface Cartesian coordinate ($\phimax$, $\thetamax$), we multiply $R(-\zeta)$ by $\emax$, 
\begin{eqnarray}
&\hat{S}&(\Phi) \, \emax^\prime \equiv R(-\zeta) \, \emax. \\
\label{eq:emaxprime}
&=& \frac{1}{L} \left(
\begin{array}{c}
\cos{(\Theta-\ThetaM)} + \cos{\ThetaM} \sin{i} \\
\cos{\zeta} [\sin{(\Theta-\ThetaM)} - \sin{i} \sin{\ThetaM}] - \cos{i} \sin{\zeta}\\
\cos{i} \cos{\zeta} + \sin{\zeta} (\sin{(\Theta-\ThetaM)} - \sin{i} \sin{\ThetaM}) 
\end{array} \right), \nonumber \\
\end{eqnarray}
Dividing the $y$-component by the $x$-component, we obtain the tangent of $\phimax + \Phi$, 
\begin{eqnarray}
\label{eq:kappa}
&\,&\tan{(\phimax + \Phi)} \nonumber \\
&=& \frac{\cos{\zeta} [\sin{(\Theta-\ThetaM)} - \sin{\ThetaM} \sin{i}] - \sin{\zeta} \cos{i}}{\cos{(\Theta-\ThetaM)}+\cos{\ThetaM} \sin{i}}. 
\end{eqnarray}

Substituting Equation (\ref{eq:kappa}) and its derivative for $\kappa(\Theta)$ and $\kappa^\prime (\Theta)$ into Equation(\ref{eq:maxfd}), we obtain the analytic forms of the modulation factor and the instantaneous frequency of the photometric variation in Equation (\ref{eq:maxf}). The explicit form of Equation (\ref{eq:maxfd}) is given in Appendix A. Figure \ref{fig:modfac} displays several examples of different geometry of the modulation factor of the maximum weighted longitude approximation. The green points indicate the singular points where the unit vector of the maximum weighted point is aligned with the spin vector $|\emax^\prime| = |\espin^\prime|$ (see Appendix A for the derivation). On the singular point, the kernel of the scattered light $W_I(\phi,\theta) W_V(\phi,\theta)$ is over the spin vector, corresponding to the maximum point of the kernel width in Figure 3 (left) of \citet{2015arXiv151105152S}. 

We can classify the instantaneous frequency curve into three domains; (A) $\zeta \le \tilde{\zeta}_I$, (B) $\tilde{\zeta}_I <  \zeta < \tilde{\zeta}_{II}$, and (C) $\zeta \ge \tilde{\zeta}_{II}$, where $\tilde{\zeta}_{I}$ and $\tilde{\zeta}_{II} (>\tilde{\zeta}_{I})$ are the obliquity of the two singular points. In the domains A, B, and C, $\epsilon(\Theta)$ has one negative peak, one negative peak and one positive peak, one positive peak, respectively. For instance, we obtain $\tilde{\zeta}_{I}=\pi/4$ and $\tilde{\zeta}_{II}=3\pi/4$ for a face-on orbit. Thus, the type of geometric frequency modulation is sensitive to the difference between the prograde, pole-on, and retrograde spins, roughly corresponding to the domains A, B, and C.

In Figure \ref{fig:modfac}, we only show one example of $\ThetaM$ for each orbital inclination. However, the topological structure is determined by the position of the singular points and the null line ($\epsilon(\Theta, \zeta) =0$) which connects the singular points. In Appendix A, we explain the general properties of the singular points and the null lines in detail. Using these features, one can roughly reproduce the general trend of the modulation factor for arbitrary parameters.

\section{Extracting the Instantaneous Frequency from Time-Frequency Representations}

\begin{figure}[htbp]
\begin{center}
\includegraphics[width=\linewidth]{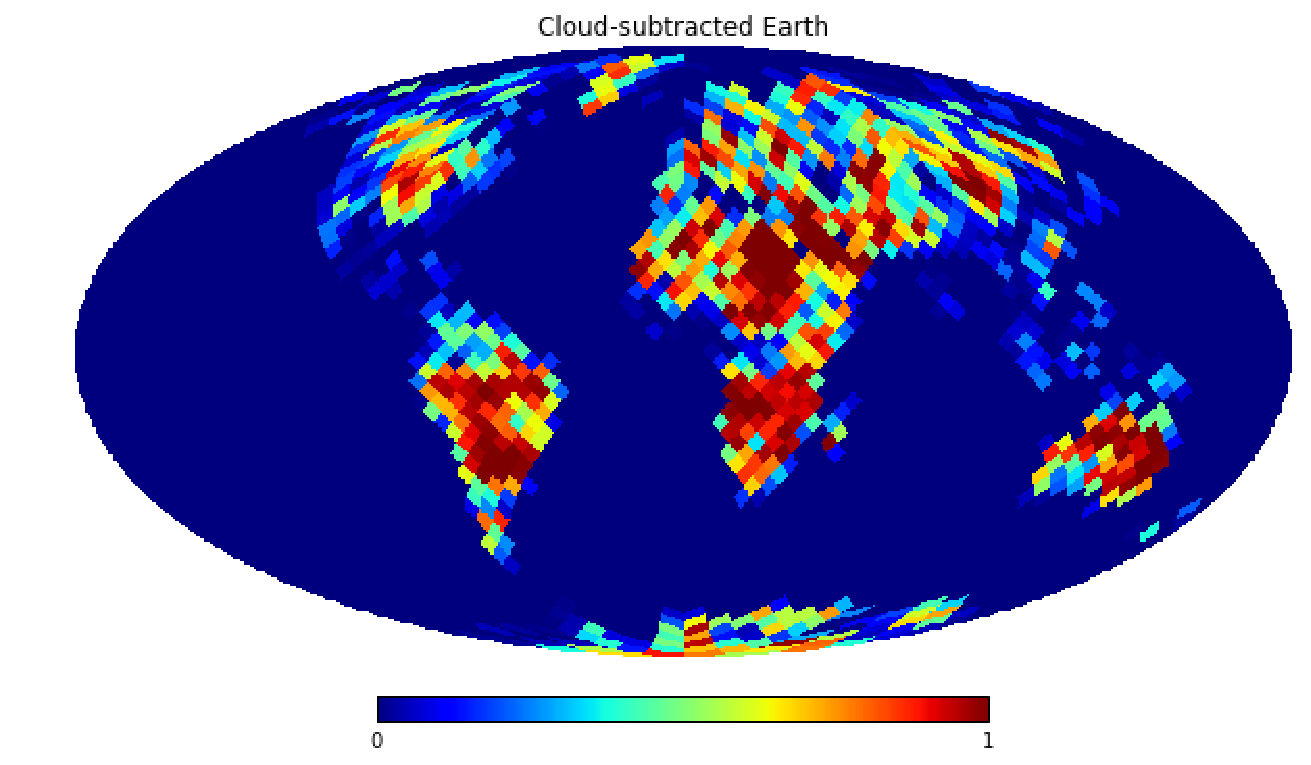}
\caption{Static cloud-subtracted Earth (static cloud-subtracted Earth) model. \label{fig:cse} }
\end{center}
\end{figure}

We compare Equation (\ref{eq:maxfd}) with the simulations of the rotational light curve whose planetary surface is stationary. We use a static cloud-subtracted Earth model. The static cloud-subtracted Earth is a toy model of the color-difference map proposed by \citet{2011ApJ...739L..62K}. In their paper, they showed that the spin-orbit tomography inferences of the obliquity using the light curve of the single band is poor because of the presence of clouds. The obliquity can be well retrieved from the color difference of the light curve (for instance 0.85 - 0.45 $\mu$ m) because the color difference of 0.85 - 0.45 $\mu$ m efficiently suppresses the effect of clouds. The static cloud-subtracted Earth model, as shown in Figure \ref{fig:cse}, has a zero-albedo ocean and a constant-albedo land after removing the cloud cover fraction. We use ISCCP D1 data (the cloud map in 2008 Jun 30 21:00) as the cloud cover fraction. The spherical pixelization is implemented using HealPix \citep{2005ApJ...622..759G} with a total pixel number of 3072.

\begin{figure}[htbp]
\begin{center}
\includegraphics[width=\linewidth]{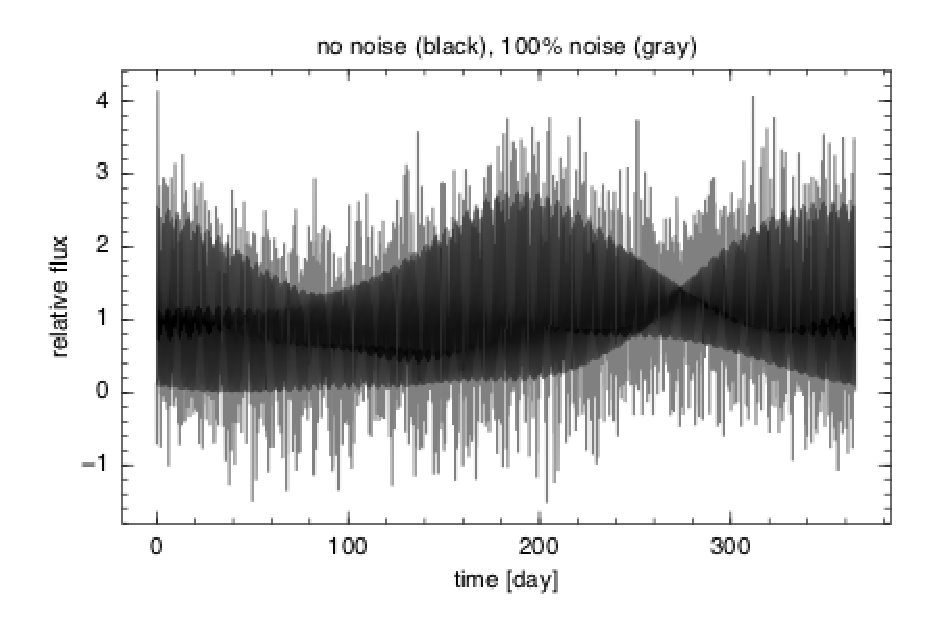}
\caption{Simulated photometric variation with no noise (black), and 100 \% noise (gray). \label{fig:sta} }
\end{center}
\end{figure}

We set $N=$4096 grids with the equal time intervals ($\sim 2.1$ hour) over a year and compute the mock photometric light curve using Equations (\ref{eq:lambert}), (\ref{eq:eS}), (\ref{eq:eO}), and, (\ref{eq:eR}). We adopt $\Pspin=23.9344699$ hr and $\Porb=365.242190402$ days corresponding to the synodic day and a year of Earth. We add Gaussian noises with the standard deviation, $\sigma_n = 0$ and $\sigma_s,$ (100\% noise) to the relative flux, where $\sigma_s$ is the standard deviation of the photometric variation. Figure \ref{fig:sta} shows examples of the generated light curve for the geometric parameter set, $\zeta = \pi/3$, $i=0$ (face-on), and $\ThetaM=\pi$. 

We estimate the instantaneous frequency by extracting the ridge of the time-frequency representation that describes the signal in both the time and frequency domains \citep[e.g.][]{cohen1995time}. We use the pseudo-Wigner distribution\footnote{The Wigner distribution, whose variable is an analytic signal of a real-valued signal, is sometimes called the Wigner-Ville distribution \citep{Boashash2015}. } as the time-frequency representation. In Appendix B, we provide a detailed description of the pseudo-Wigner distribution as an instantaneous frequency estimator. Here we provide a summary. The pseudo-Wigner distribution is expressed as 
\begin{eqnarray}
\label{eq:pseudowv}
g(\fif, t) = \int_{-\infty}^{\infty} h(\tau) z (t + \tau/2) z^\ast (t - \tau/2) e^{-2 \pi i \mathit{f} \tau } d \tau, \nonumber \\
\end{eqnarray}
where $z$ and $z^\ast$ are the analytic signal of the data and its conjugate, and $h(\tau)$ is the window. For the discrete sequence, $z[1],z[2],...z[N]$
\begin{eqnarray}
\label{eq:pseudowvd}
g(\fif, t_i) = \sum_{|m|<N/2} h[m] z [i+m] z^\ast [i-m] e^{-2 \pi i \mathit{f} \tau } d \tau, \nonumber \\
\end{eqnarray}
where $m = \mathrm{min}(i-1,N-i)$. Using Equation (\ref{eq:pseudowvd}), we compute the time-frequency representation between $f_i=0.96$ and $f_j=1.05$ [1/day] around $\mathit{f} = 1/\Pspin$. The ridge line of the time-frequency representation is interpreted as the instantaneous frequency,
\begin{eqnarray}
\label{eq:argmax}
\hat{\fif} (t) = \mathrm{argmax}_\mathit{[f_i, f_j]} g(\fif, t),
\end{eqnarray}
where $[f_i, f_j]$ is the frequency range of interest. 

We translated the MATLAB code for computing Equation (\ref{eq:pseudowvd}) in the Time-Frequency Toolbox \footnote{http://tftb.nongnu.org} to Julia language \citep{2014arXiv1411.1607B}. The fact that $\mathit{f}$ is not necessarily discrete is important for our purpose because the frequency modulation is small (of the order of $\delta \mathit{f}/\mathit{f} \approx \Pspin/\Porb$). Practically, the fast Fourier transform (FFT) is inefficient for exploring the narrow frequency range of interest. We replaced the FFT in the original code by the non-uniform FFT algorithm \citep{2004SIAMR..46..443G}. In Appendix C, we describe the implementation of the non-uniform FFT into the pseudo-Wigner distribution and the comparison of the systems with the non-uniform FFT and the FFT. We also test our code for known signals with a given instantaneous frequency in Appendices C and D. In practice, the selection of the window width of $h(\tau)$, $\mathrm{w}$, is critical for noisy data. We adopt a window width of $\mathrm{w}=N/8$ for the 100 \% noise. Our code used in the paper is publicly available under the GNU General Public Licence. \footnote{https://github.com/HajimeKawahara/juwvid}

The analytic signal $z[1],z[2],...,z[N]$ in Equation (\ref{eq:pseudowvd}) is generated using the following procedure. To remove the amplitude modulation (detrending of the amplitude), we compute the interpolation function of the mock photometric light curve of the mean and standard deviation for each set of 64 adjacent data points ($\sim $ six days). We subtract the interpolated mean from the light curve and divide by the interpolated standard deviation. This pre-processing enables us to exclude unnecessary amplitude modulations from the frequency analysis and normalizes the light curve. Then, we compute the analytic signal of the normalized light curve using the DSP package of Julia. 

\begin{figure}[htbp]
\begin{center}
\includegraphics[width=\linewidth]{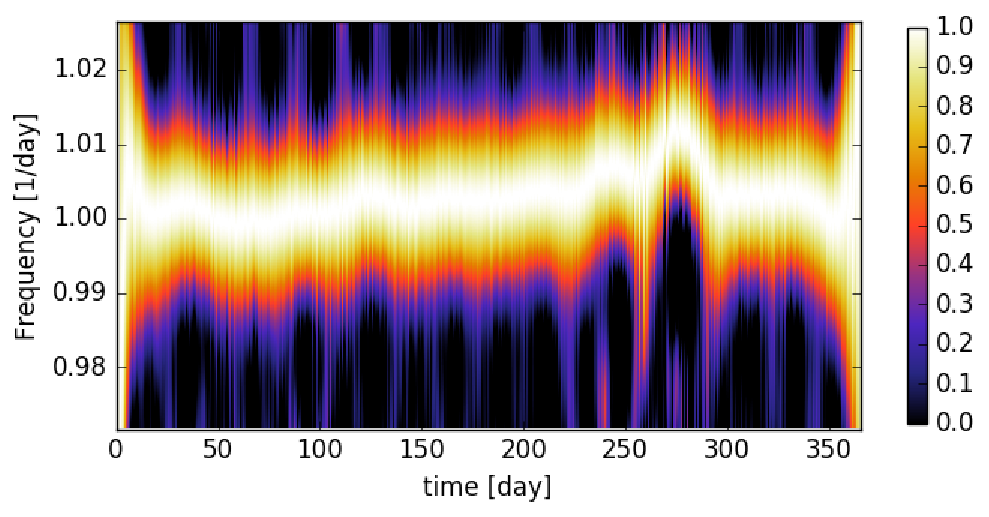}
\caption{Time-frequency representation by the pseudo-Wigner distribution of the light curve with the 100 \% noise \label{fig:tfr} }
\end{center}
\end{figure}

\begin{figure}[htbp]
\begin{center}
\includegraphics[width=\linewidth]{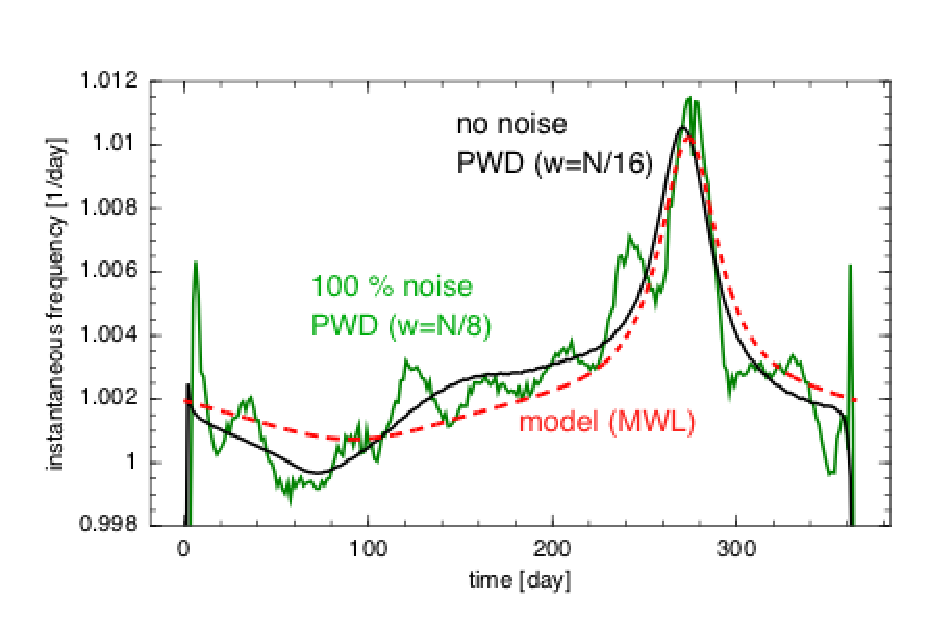}
\caption{Estimated instantaneous frequencies by the pseudo-Wigner distribution (labeled as PWD) for the light curves with no noise (black) and 100 \% noise (green). The dashed curve (red) is the theoretical prediction from the maximum weighted longitude approximation (labeled as MWL) \label{fig:if1} }
\end{center}
\end{figure}

Figure \ref{fig:tfr} shows an example of the time-frequency representation of the 100 \% noise case for $\zeta=60^\circ, \ThetaM=180^\circ, i=0^\circ$. Extracting the ridge line of the time-frequency representation, we extract the extracted instantaneous frequency shown by the green line in Figure \ref{fig:if1}. The black curve in Figure \ref{fig:if1} is the extracted instantaneous frequencies of the light curve with no noise ($\mathrm{w}=N/16$). The extracted instantaneous frequency of the 100 \% noise still exhibits the same characteristic features of the geometric effect as those of of the noiseless case. The dashed line indicates the instantaneous frequency from the maximum weighted longitude approximation. The prediction curve from the maximum weighted longitude approximation reproduces the general characteristics of the instantaneous frequency of the simulations. However, there remains some difference between the black and red lines. Because the test given in Appendix C does not exhibit such a difference (Figure \ref{fig:width}), this difference originates from the influence of the albedo distribution we ignore in the maximum weighted longitude approximation. 

We investigate another dozen parameter sets for $\zeta, \ThetaM$. We confirmed that we can extract the instantaneous frequency for the $\sim$ 100 \% noise in most cases but, in some configurations, we need to decrease the noise level to $\sim 30 $\% to extract the instantaneous frequency. We also found that similar cases sometimes happen when randomly rotating the albedo distribution for the fiducial parameter set.  We found that these noise-sensitive cases can be attributed to particular configurations which prevent sufficient amplitude of the photometric variation depending on the albedo distribution and the geometric parameters. Since optimizing the noise is beyond the scope of our paper, we do not try to improve this further to detect the instantaneous frequency for these noise-sensitive cases. Instead, we point out that numerous techniques in the high-noise environment have been proposed \citep[see ][]{stankovic2013time,Boashash2015}. These techniques may potentially improve the detectability of the noise-sensitive configurations.

\subsection{Comparison with the maximum weighted longitude model}

\begin{figure*}[htbp]
\begin{center}
\includegraphics[width=0.32\linewidth]{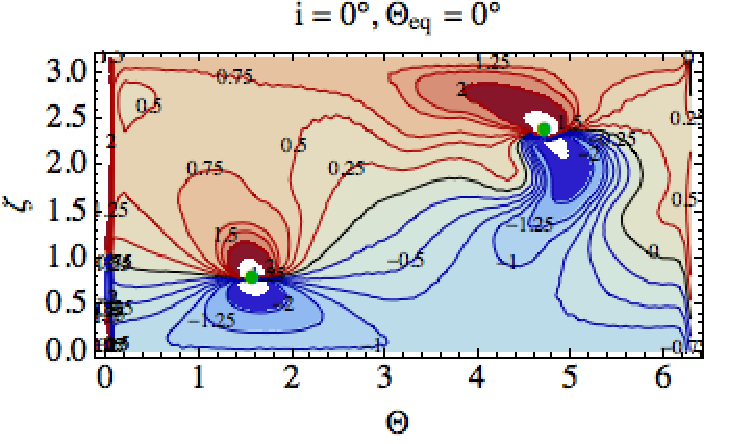}
\includegraphics[width=0.32\linewidth]{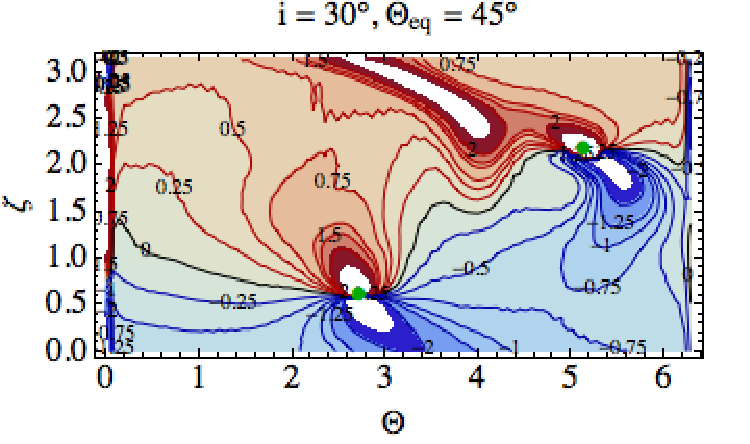}
\includegraphics[width=0.32\linewidth]{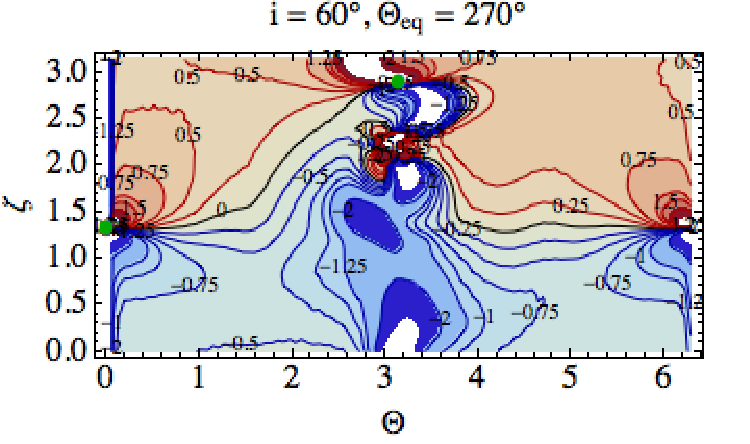}
\caption{\label{fig:modfacsim} Examples of the simulated modulation factor using the pseudo-Winger distribution, corresponding to the panels in Figure \ref{fig:modfac}. }
\end{center}
\end{figure*}

Figure \ref{fig:modfacsim} displays the modulation factor of the simulated light curve derived from the pseudo-Wigner distribution. We used the light curves with no noise, and adopted the pseudo-Wigner distribution with a window width of $\mathrm{w}=N/16$ to suppress the bias (see Appendix D). Compared with Figure \ref{fig:modfac}, we find that the maximum weighted longitude approximation provides close agreement with the simulated modulation factor despite the fact that we ignore the dependence of the albedo distribution of the frequency modulation. Thus, the frequency modulation contains sufficient information on the obliquity. 

For the highly inclined case $(i \gtrsim 60^\circ)$, the intensity of the light curve significantly decreases near the inferior conjunction, and the gradient of the light curve also increases. These disadvantages near the inferior conjunction prevent accurate measurements of the frequency modulation due to the orbital inclination. Even for $i=60^\circ$ (the right panel in Figure \ref{fig:modfacsim}), one can see the disturbance of the extracted instantaneous frequency near the inferior conjunction $\Theta = \pi$. However, the difficulty for highly inclined planets is less problematic for our purpose because the direct imaging at the inferior conjunction for highly inclined planets is challenging in the first place. 

To use the instantaneous frequency as an obliquity estimator, we fit the extracted instantaneous frequency by the maximum weighted longitude model. We generate 1000 realizations of the light curves with the noise for each parameter set and extract the instantaneous frequencies with the pseudo-Wigner distribution ($\mathrm{w}=N/8$). We fit them by the maximum weighted longitude approximation using the Levenberg-Marquardt algorithm \citep[mpfit][]{2009ASPC..411..251M}. We regard $\Porb$ and $i$ as known parameters because the monitoring observation of the direct imaging provides them. Then, there remain the there free parameters $\zeta$, $\ThetaM$, and the spin rotation period. To avoid the aliasing effect, we exclude the most left and most right $\sim$ 19 days of the extracted instantaneous frequency. 

\begin{figure}[htbp]
\begin{center}
\includegraphics[width=\linewidth]{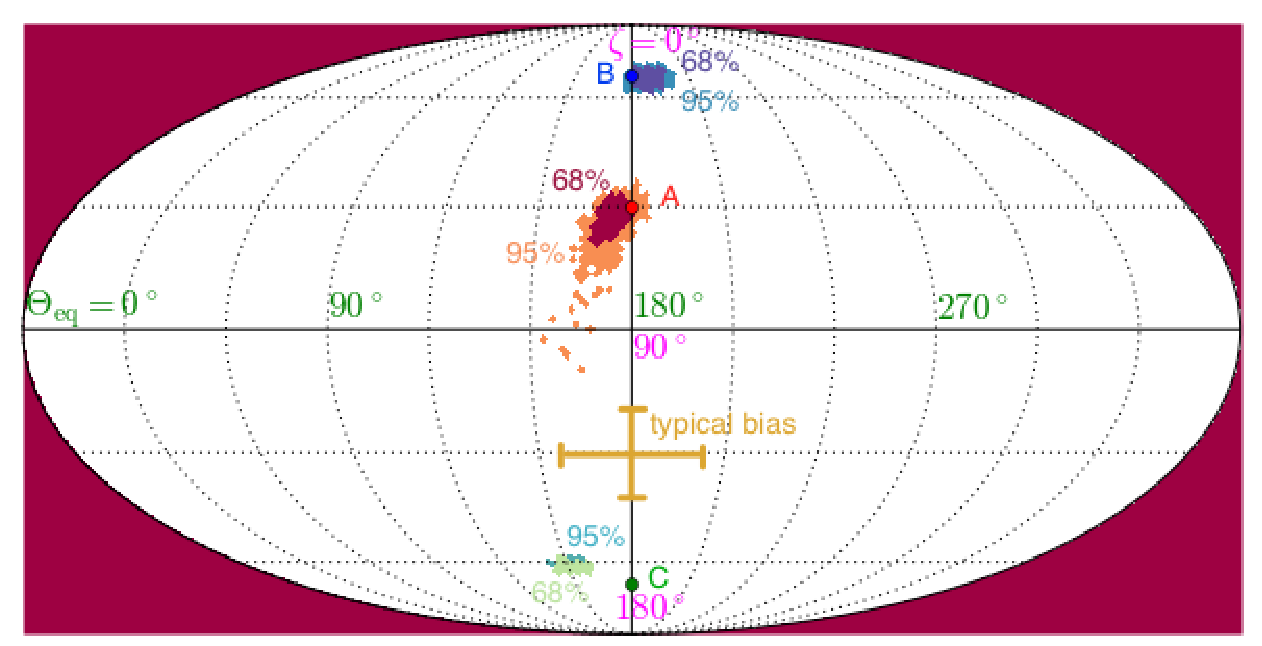}
\includegraphics[width=\linewidth]{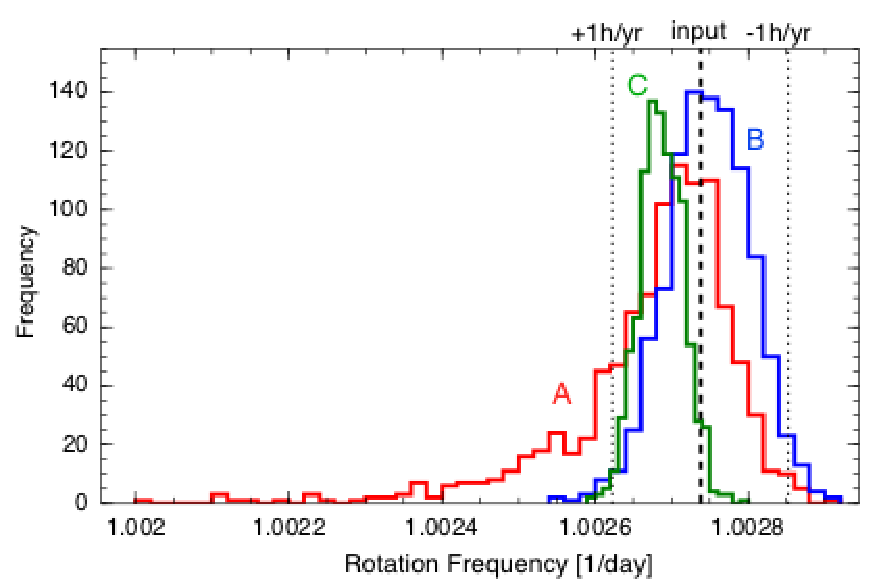}
\caption{Areas enclosing 68 \% and 90 \% of the best-fit $\zeta$ and $\ThetaM$ (top), and a histogram of the best-fit spin rotation frequency (bottom) for the sets of the 1000 realizations. We consider three cases: (A) $\zeta=60^\circ$ (red), (B) $\zeta=23^\circ$ (blue), and (C) $\zeta=157^\circ$ (green). In the top panel, the input value is marked by the points. The yellow cross indicates the typical bias discussed in the text. In the bottom panel, the vertical dashed line and the dotted lines indicate the input value and the one hour errors in one year $\Delta \fspin \sim 1/(24 \times 365)$ [1/day].  \label{fig:est}}
\end{center}
\end{figure}

Figure \ref{fig:est} (top) shows the areas enclosing 68 \% and 95 \% of the best-fit obliquity and phase. We assume a face-on orbit and $\ThetaM=180^\circ$ and test three different obliquities: (A) $\zeta=60^\circ$, (B) $\zeta=23^\circ$, and (C)$\zeta=157^\circ$. For cases A and B, we set noise at 100 \%, but we reduce the noise to 30 \% for the case C because it has the noise-sensitive configuration. The points mark the input values. The estimated planetary obliquity and the estimated phase are in good agreement with the input values for the cases A and B, although there are slight shifts of the center from the input. For case C, the bias is much larger than the statistical uncertainty. This bias is due to the lack of the albedo information in the maximum weighted longitude approximation. Although the maximum weighted longitude approximation (Figure \ref{fig:modfac}) roughly reproduces the simulated modulation factor (Figure \ref{fig:modfacsim}), the peak positions for a given $\zeta$ between Figures \ref{fig:modfac} and \ref{fig:modfacsim} often differs by several tens of degrees in $\ThetaM$. This difference also impacts the uncertainty of the obliquity. Considering these facts, the confidence region of the fitting with the maximum weighted longitude approximation is typically $\Delta \zeta  \sim 20^\circ$ and $\Delta \ThetaM  \sim 45^\circ$, as shown by the yellow cross. To further reduce the bias, one requires the modeling of the albedo distribution in the instantaneous frequency model. We do not consider albedo modeling in this paper.

The uncertainty of the spin rotation frequency relies on the determination of the offset of the instantaneous frequency. As shown in the bottom panel of Figure \ref{fig:est}, the best-fit spin rotation frequency agrees with the input one with $\sim 0.02-0.03$ \% accuracy, corresponding to a few hours of error during one year. 

\section{Discussion}

\subsection{Light travel time effect}

The finite light-travel time effect, known as R{\o}mer delay, also induces the frequency modulation. The light-travel time effect provides direct information on the length of the system. The light-travel time effect has been detected in the forms of the modulation of the eclipse timing of the hierarchical triple-star systems \citep{2013ApJ...768...33R,2014AJ....147...45C,2015ApJ...806L..37M,2016MNRAS.455.4136B} and the time delay between the transit and secondary eclipse of exoplanets \citep{2010ApJ...721.1861A}. Pulsating stars are also used as a clock to measure the frequency modulation \citep{2012MNRAS.422..738S}. Here, we consider the light-travel time effect for a directly imaged planet in a circular orbit. The time delay is $\Delta t = - a \cos{\Theta} \sin{i}/c$, where $a$ and $c$ are the semimajor axis and the speed of light. The phase of the diurnal periodicity is given by 
\begin{eqnarray}
\label{eq:phased}
\psi(t) &=& 2 \pi \fspin ( t + \Delta t) + \mathrm{const}.
\end{eqnarray}
The instantaneous frequency is expressed as
\begin{eqnarray}
\label{eq:phasedif}
\fobs &=& \frac{1}{2 \pi} \frac{\partial \psi}{\partial t} = \fspin + \epsilon_{LT}(\Theta) \forb \\
\epsilon_{LT}(\Theta) &=& 2 \pi \left(\frac{\fspin a}{c}\right) \sin{i} \sin{\Theta} \\ 
&=& 0.036 \left(\frac{a}{\mathrm{au}} \right) \left(\frac{\fspin}{\mathrm{day^{-1}}} \right)  \sin{i} \sin{\Theta}. 
\end{eqnarray}
The amplitude of the frequency modulation by the light-travel time effect $\Delta f = |\fobs - \fspin|$ is of the order of $10^{-4}$ [1/day] for $a=1$ au, $\fspin=1$ [1/day], $\forb=1/365$ [1/day]. Thus, the frequency modulation due to the light-travel time effect is two orders of magnitude smaller than the geometric effect with $a=1$ au and a similar spin rotation rate to that of Earth. Because the thermal light also has the light-travel time effect, the light-travel time effect becomes important for thermal emission from a long-period planet with $a > 10$ au, instead of the scattered light.  

\subsection{Planetary Wind}

The geometric effect dominates the frequency modulation for atmosphereless planets. Indeed, atmosphereless planets can exhibit significant photometric variation \citep{2014AsBio..14..753F}. For thin-atmosphere planets like Earth, the dynamic effect of clouds is important. The static cloud-subtracted Earth model we used in this paper implicitly assumes that the effect of clouds can be efficiently removed by the difference of two bands \citep{2011ApJ...739L..62K}. Principal component analysis of the multi-band observation will help to make a proper combination of the bands to eliminate clouds \citep{2013ApJ...765L..17C}. The validity of the assumption of the static component is of particular importance for discussing the feasibility because the seasonal changes of the global planetary winds can be a possible source of frequency modulation. 

Analyzing realistic simulations of a single-band observation of a mock Earth, \citet{2008ApJ...676.1319P} reported slight shifts in the best-fit rotational period to shorter periods. They concluded that the variable cloud cover produced the shifts. The order of the geometric frequency modulation for Earth is $\forb/\fspin$ = 0.3 \%, corresponding to a wind speed of 1.3 $\mathrm{m s^{-1}}$. The seasonal change or global difference of the wind velocity $\lesssim 1 \mathrm{m s^{-1}}$  can be comparable to the geometric effect. The possibility of the detection of planetary wind is an interesting topic in itself, and the model of geometric frequency modulation should still be critical to extract the modulation of the planetary winds. To investigate the effect of the global planetary wind, detailed simulations of radiative transfer with a sufficiently fine time resolution are required. We postpone the impact of the dynamic planetary surface to a forthcoming paper using a global climate model and satellite data (Kawahara and Kodama in preparation).  

Application of our results to gas giants requires that we consider differential rotation. For instance, the zonal wind ($v \sim 100$ km/s) in Jupiter reaches $\sim 1$ \% of the spin rotation \citep{2005RPPh...68.1935V}, which is one order of magnitude larger than $\forb/\fspin$ even if we assume that this Jupiter is located at 1 au. In this case, the differential rotation significantly affects the modulation factor. The assumption of a static surface is no longer valid.

\subsection{Comparison with inversion techniques based on the amplitude modulation}

It is worth comparing our method with the traditional inversion, which uses the amplitude modulation \citep[AM-based method; ][]{2010ApJ...720.1333K, 2015arXiv151105152S}. Our method relies on the zonal inhomogeneity of the planet because the change of the longitudinal location generates the frequency modulation. In the AM-based method, \citet{2015arXiv151105152S} showed that the longitudinal width (not the longitudinal location) and the dominant colatitude constrain the obliquity. Although the AM-based method and our techniques utilize different properties of the longitudinal kernel (i.e. width versus location), a narrow longitudinal width of the kernel increases both the amplitude and the frequency modulation. In this sense, the two methods extract the same information on the kernel position in the different ways.

Our approach is also complementary to the AM-based technique even for practical reasons. \citet{2015arXiv151105152S} showed that one could in principle identify the obliquity of a planet with high-precision observations of a single rotation at only two orbital phases. Our method requires full-orbit observations, but works for noisy data. 

The amplitude modulation does not provide a map-independent scheme for extracting planetary obliquity. Therefore, the AM-based method suffers from a severe albedo-radius degeneracy. Our method does not require that we solve the albedo map of the planet. The frequency modulation technique sidesteps this problem. However, our method requires that the albedo map be static, whereas the AM-based method can in principle work for time-variable maps \citep{2015arXiv151105152S}. \citet{2011ApJ...739L..62K} and \citet{2012ApJ...755..101F} showed that the spin-orbit tomography may constrain the obliquity for the simulated Earth with cloud variability. Similarly, further simulations including the cloud variability are required to validate the frequency modulation technique for Earth-like planets.

The original spin-orbit tomography does not distinguish prograde rotation from retrograde one \citep[i.e. $\zeta$ from $\pi - \zeta$; ][]{2010ApJ...720.1333K}. \citet{2015arXiv151105152S} suggested that the longitudinal location of the kernel over an orbit might differentiate between the prograde and retrograde rotation. Frequency modulation is very sensitive to the difference between the prograde and retrograde rotation, even for a face-on orbit, as mentioned at the end of Section 2. 

Another advantage of using the frequency domain is that information on the normalization is not required. As is apparent from the procedure for the amplitude detrending, long-term stability over a year is not significant. Moreover, the dependence of the geometric parameters on the instantaneous frequency is intuitively comprehensible. Regarding the statistical noise, we showed that the pseudo-Wigner distribution retrieved the instantaneous frequency of the data with the 100\% noise compared to the standard deviation of the signal (for 2-hr). Because the typical amplitude of the variation of Earth is $\sim 10$ \%, it corresponds to S/N $\sim$ 10 for a 2 hr exposure. \citet{2012ApJ...755..101F} assume the S/N=20 for a 4.8 hr exposure for the spin-orbit tomography, corresponding to the photon limit of a 5 m telescope for a 10 pc Earth. Our considered S/N is comparable to theirs.

The full inversion by the spin-orbit tomography requires a precise measurement of the spin rotation period. For instance, one can recognize the frequency modulation of the simulated light curves in the diurnal and orbital phase plane in Figure 6 of \citet{2010ApJ...720.1333K}, which were used for the retrieval of the surface map and obliquity. However, to obtain such images, we require the precise value of the spin frequency with uncertainty below $\sim 0.2 \Pspin/\Porb$ ($\sim$ 5 hr error in one year). It has not been shown how to measure the spin rotation with such precision. The time-frequency analysis can provide the precise value with an uncertainty of the order of $0.1 \Pspin/\Porb$. which is critical even for the spin-orbit tomography.

\section{Summary}

In this paper, we found that the axial tilt and the orbital inclination induce the frequency modulation of the apparent periodicity of the scattered light. We constructed the analytic model of the instantaneous frequency, which has three geometric parameters: the obliquity, the seasonal phase, and the spin rotation period. Fitting the instantaneous frequency extracted from the pseudo-Wigner distribution of the simulated light curve, we demonstrated that one can infer these parameters from the time-frequency analysis of the light curve of directly imaged planets. The frequency modulation provides a complementary technique to the inversion based on the amplitude modulation. 

H.K. is supported by a Grant-in-Aid for Young Scientists (B) from the Japan Society for Promotion of Science (JSPS), No.\,25800106. We are grateful to the referee, Nick Cowan, for many helpful suggestions, in particular, an insightful consideration of the comparison with the AM-based method. 

\appendix

\section{A. Analytic expressions}

The modulation factor on the maximum weighted longitude approximation is expressed as 
\begin{eqnarray}
\epsilon_\zeta(\Theta) = \frac{-\cos{\zeta} + \sin{\zeta} \cos{i} \sin {\Delta \Theta} - \cos{\zeta} \cos{\Theta} \sin{i}}{\cos ^2{\Delta \Theta}+(-\cos{\zeta} \sin{\Delta \Theta}+\cos{\zeta} \sin{\ThetaM} \sin{i}+\sin{\zeta} \cos{i})^2+2 \cos{\ThetaM} \sin{i} \cos{\Delta \Theta}+\cos^2{\ThetaM} \sin ^2{i}}, 
\end{eqnarray}
where $\Delta \Theta \equiv \Theta -\ThetaM$.

The singular point satisfies that $|\emax^\prime| = |\espin^\prime|$. In other words, the x- and y- components of Equation (\ref{eq:emaxprime}) should be zero. Then, we obtain the singular point $(\tilde{\zeta},\tilde{\Theta})$ for $\ThetaM = \tThetaM$
\begin{eqnarray}
\cos{(\tilde{\Theta} - \tThetaM)} &=& - \sin{i} \cos{\tThetaM} \\
\tan{\tilde{\zeta}} &=& \tan{i} ( - \sin{\tThetaM} \pm \sqrt{\csc^2{i} - \cos^2{\tThetaM}} ).
\end{eqnarray}
The parameter set that satisfies the above equations corresponds to the singular point where $\phimax$ cannot be defined.

\begin{figure}[htbp]
\begin{center}
\includegraphics[width=0.49\linewidth]{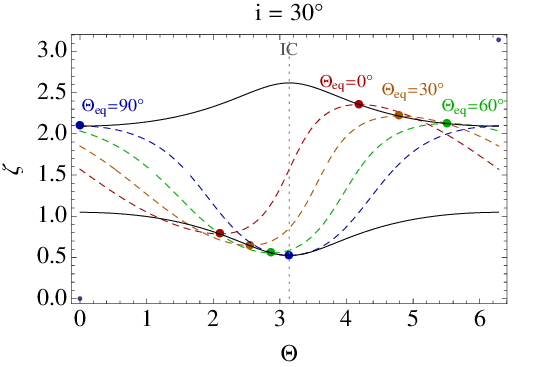}
\includegraphics[width=0.49\linewidth]{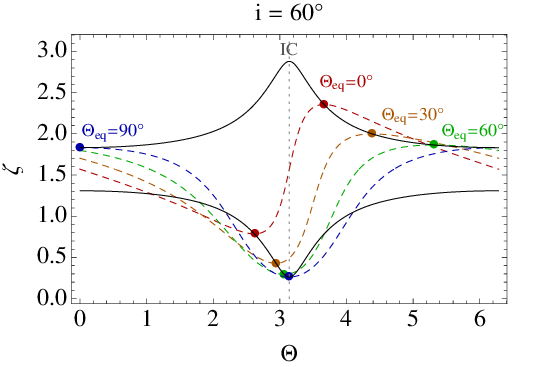}
\caption{The singular points (marked by filled circles) and the null line ($\epsilon(\zeta, \ThetaM)=0$) as a function of $\ThetaM$ (colors) for $i=30^\circ$ and $60^\circ$. The black solid lines are trajectory of the singular point. \label{fig:spnl} }
\end{center}
\end{figure}

The dependence of the singular points on $\ThetaM$ is shown in Figure \ref{fig:spnl}. As $\ThetaM$ changes, the singular point primarily runs parallel to $\Theta$. However, when one of the singular points is passing through the inferior conjunction (denoted by IC), the inclination effect traps the singular point. Then, the interval between the pair of the singular points becomes narrow. As the orbital inclination increases, the trapping effect becomes stronger. We also plot the null lines on which the modulation factor is zero:
\begin{eqnarray}
\tan{\zeta_\mathrm{null}} = \frac{1 + \cos{\Theta} \sin{i}}{\cos{i} \sin{(\Theta - \ThetaM)}}.
\end{eqnarray}
Thus, one can roughly imagine and understand the general feature of the modulation factor from the position of the singular points and the null line. 

\section{B. The instantaneous frequency extraction using the Wigner Distribution}

Following \citet{cohen1995time, stankovic2013time, Boashash2015}, we briefly summarize the Wigner distribution and the pseudo-Wigner distribution as the instantaneous frequency estimator. Let us consider the normalized analytic signal $z(t) = e^{i \psi(t)}$, where $\psi(t)$ is the instantaneous phase. The instantaneous frequency is defined by 
\begin{eqnarray}
\fif (t) = \frac{1}{2 \pi} \frac{\partial \psi (t)}{\partial t}.
\end{eqnarray}
The ideal time-frequency representation should be expressed as 
\begin{eqnarray}
\rho (\fif, t) \propto \delta_D (\fif - \mathit{\hat{f}}(t)),
\end{eqnarray}
where $\delta_D (x)$ is the delta function and $\hat{f}(t)$ is the instantaneous frequency. The inverse fourier transform ($\fif \to \tau$) is 
\begin{eqnarray}
\label{eq:hatrho}
\tilde{\rho} (\tau, t) = e^{2 \pi i \mathit{\hat{f}}(t) \tau} = \exp{\left( i \tau \frac{\partial \psi (t)}{\partial t}\right)}.
\end{eqnarray}
Approximating the derivative of the instantaneous phase by 
\begin{eqnarray}
 \frac{\partial \psi (t)}{\partial t} \approx  \frac{\psi (t + \tau/2) - \psi (t - \tau/2)}{\tau},
\end{eqnarray}
for a small time step $\tau$, one obtains the Wigner distribution from the Fourier transform of the approximated $\hat{\rho} (\fif, \tau)$, 
\begin{eqnarray}
\rho (\fif, t) &=& \int_{-\infty}^\infty \tilde{\rho} (\tau,t) e^{- 2 \pi i \fif \tau} d \tau \\
&\approx& \int_{-\infty}^\infty \exp{\left[ i \psi (t + \tau/2) - i \psi (t - \tau/2) \right]} e^{- 2 \pi i \fif \tau} d \tau \\
&=& \int_{-\infty}^\infty z(t+\tau/2) z^\ast(t-\tau/2) e^{- 2 \pi i \fif \tau} d \tau 
\end{eqnarray}

The pseudo-Wigner distribution is the windowed version of the Wigner distribution, which emphasizes the properties near the time of interest $t$ and suppresses the cross-term of the noise \citep{cohen1995time},
\begin{eqnarray}
\label{eq:pseudowvax}
g(\fif, t) = \int_{-\infty}^{\infty} h(\tau) z (t + \tau/2) z^\ast (t - \tau/2) e^{-2 \pi i \mathit{f} \tau } d \tau. 
\end{eqnarray}
Because the pseudo-Wigner distribution can be expressed as the convolution of the Wigner distribution and the fourier conjugate of the window,  
\begin{eqnarray}
\label{eq:pseudowvsax}
g(\fif, t) = \tilde{h} \ast \rho (\fif, t),
\end{eqnarray}
the pseudo-Wigner distribution is the smoothed version of the Wigner distribution in the frequency domain \citep{stankovic2013time}. The Hamming window is given by 
\begin{eqnarray}
h(\tau) &=& 0.54 + 0.46 \cos{\left(2 \pi \frac{\tau}{\omega} \right)} \mbox{\,\,\,\,\, for $|\tau| \le \omega/2$} \nonumber \\
 &=& 0 \mbox{\,\,\,\,\, otherwise} \\
\end{eqnarray}
where $\omega$ is the window width. 

\section{C. Testing the pseudo-Wigner distribution using the non-uniform FFT}

The discrete pseudo distribution is expressed as 
\begin{eqnarray}
\label{eq:pseudowvda}
g(\fif, t_i) = \sum_{|m|<N/2} h[m] z [i+m] z^\ast [i-m] e^{-2 \pi i \mathit{f} \tau } d \tau,
\end{eqnarray}
where $N$ is the number of data. The ridge line of the pseudo-Wigner distribution is interpreted as the instantaneous frequency,
\begin{eqnarray}
\label{eq:argmaxa}
\hat{\fif} (t) = \mathrm{argmax}_\mathit{[f_i, f_j]} g(\fif, t),
\end{eqnarray}
where $[f_i, f_j]$ is the frequency range of interest.

To test our code, we generate mock data sets for a given instantaneous frequency function of $\hat{\fif}(t)$,
\begin{eqnarray}
\label{eq:cum}
y_j &=& \cos{\left( \frac{t_N}{N} \sum_{i=1}^{j} \frac{\hat{\fif}(t_i)}{2 \pi} \right)}.
\end{eqnarray}
We adopt $N=4096$ and the model of the maximum weighted longitude approximation used in Figure \ref{fig:if1} as $\hat{\fif}(t)$. 

\begin{figure}[htbp]
\begin{center}
\includegraphics[width=0.49\linewidth]{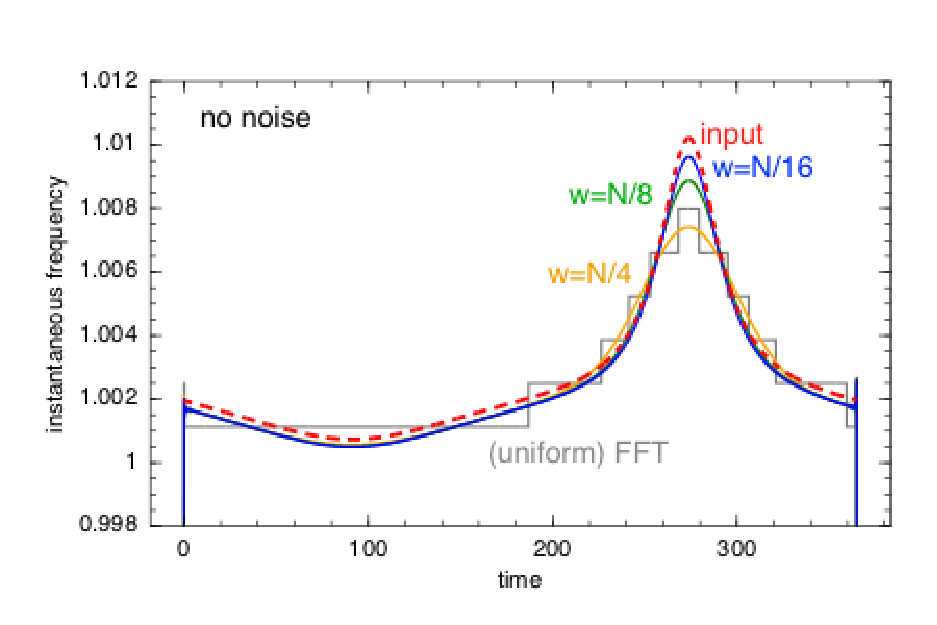}
\includegraphics[width=0.49\linewidth]{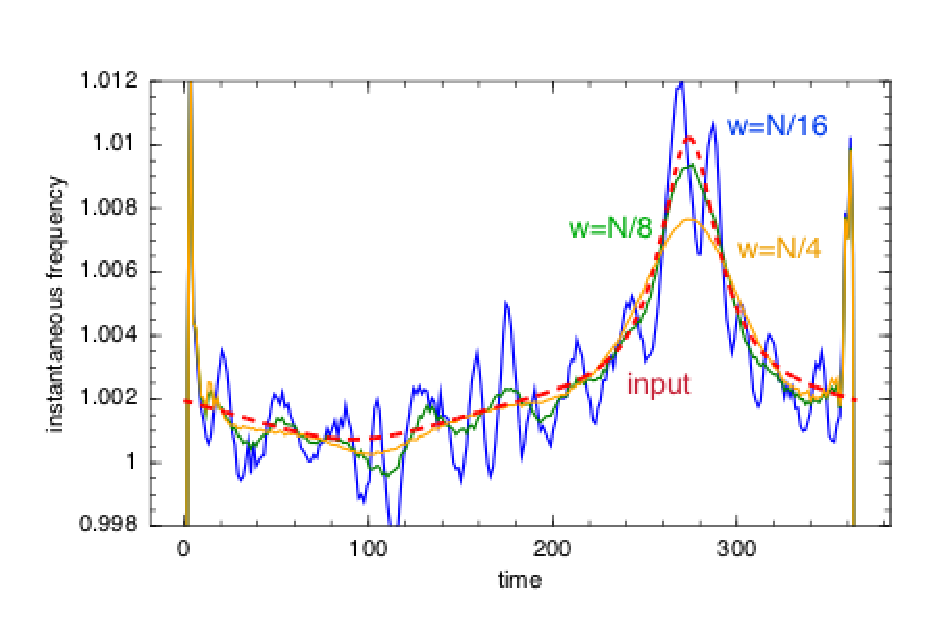}
\caption{Left: A test of the instantaneous frequency estimation from the pseudo-Wigner distribution. The gray line uses the FFT as a Fourier-transform solver. The blue, green, and yellow curves use the non-uniform FFT with the window width of $\mathrm{w}=N/16, N/8$, and $N/4$. The dashed curve indicates the input instantaneous frequency. Right: the extracted instantaneous frequency for the data with the 50 \% noise. The curves are same to the left panel.  \label{fig:width} }
\end{center}
\end{figure}

As is apparent from equation (\ref{eq:pseudowvda}), the frequency is an arbitrary value (not a discrete value) under the Nyquist frequency. If using the FFT to solve equation (\ref{eq:pseudowvda}), the sampling rate of the frequency is $1/\Delta t$. The gray line in Figure \ref{fig:width} displays an example of the extracted instantaneous frequency using the FFT. The coarse sampling rate is due to the uniform grid. The non-uniform FFT enables us to increase the sampling rate efficiently in $O(n \log{n})$ operations \citep{2004SIAMR..46..443G}. The other solid lines in Figure \ref{fig:width} indicate the extracted instantaneous frequency  using the non-uniform FFT with 4096 grids between $f=0.972$ and $1.027$ [1/day].The computational costs of the codes with the FFT and the non-uniform FFT are on the same order. 

\section{D. Dependence of the window width on the bias for the nonlinear instantaneous frequency and the noise suppression}

The selection of the window width affects both the bias in the frequency direction and the suppression of the noise. The blue, green, and yellow curves in the left panel of Figure \ref{fig:width} correspond to the extracted instantaneous frequency using $\mathrm{w}=N/16, N/8$ and $N/4$, where $\mathrm{w} = \omega N/t_N$. In the left panel, we do not add any additional noise. As the window size decreases, the bias at a nonlinear instantaneous frequency point becomes smaller. However, there is a trade-off relation between the frequency bias and the noise suppression (or frequency resolution). Figure \ref{fig:width} shows the extracted instantaneous frequency using the same window sizes as in the left panel for the data with additional noises (50 \% of the standard deviation of the signal). The extracted instantaneous frequency with a smaller window exhibits higher noises (or poor resolution of the instantaneous frequency). Thus, the adequate size of the window depends on the noise level of the data. 

For the highly nonlinear instantaneous frequency, the adaptive algorithm that determines the appropriate window size as a function of time was proposed \citep[Section 5 of ][]{stankovic2013time}. We tried to use the adaptive algorithm and found that the adaptive algorithm chooses a smaller window at the peak of the instantaneous frequency. However, there are other artifacts from the adaptive algorithm, and our data is not likely to be highly nonlinear. Therefore, we decide to use a constant window size for simplicity in this paper.


\end{document}